\documentclass[review]{elsarticle}
\begin{document}

\begin{frontmatter}

\title{COHERENT DIBARYONS}

\author{B.F. Kostenko}
\address{Joint Institute for Nuclear Research \\ Joliot-Curie 6,
Dubna, Moscow region, Russia\\
e-mail:bkostenko@jinr.ru}

\begin{abstract}
  There are encouraging  evidences for registration of excitations of
quark oscillator levels  in compressed 2-nucleon systems, which were
hidden in two independently published experimental works for many
years. Data obtained by EVA collaboration can be considered as the
third possible confirmation of this conjecture. Moreover, they drop
hints about existence of coherent dibaryons which may be interpreted
as usual or generalized coherent  states of the oscillator. These
data may also be an explicit example of a physical process described
by Landau and Peierls in their reasoning concerning the energy-time
uncertainty relation. Two models of the coherent dibaryon creation
are formulated. Phase transitions of baryon matter in few-nucleon
systems are discussed.
\end{abstract}

\begin{keyword}
 Resonances \sep two-nucleon system \sep quark matter
  \sep phase transitions
\end{keyword}

\end{frontmatter}


\section{Introduction}
It was suggested in \cite{Kostenko1, Kostenko3} to use a
cumulative particle (CP) as a trigger for a detection of the
multibaryon (MB) production. Our estimations showed that appearance
of CP is a signature of {\it ``deep cooling"} of a multiquark system,
which brings it close to its ground state and allows it to have a
narrow width thereby giving a chance to separate MB from a secondary
particle background. Another method of deep cooling was described
implicitly in \cite{Troyan}. In this case, events of n-p
interactions with {\it sufficient numbers} of secondary pions  were
selected and dibaryons were supposed to be observed as peaks on the
plot of yield versus effective masses of outgoing p-p pairs. These
data may be described by a simple formula $M_{2B}  \approx M_{pp}+10
n$ MeV, where $n$ is a positive integer. Although not each such $n$
was matched with a dibaryon in \cite{Troyan}, it was natural to
assume that they exist too but were not seen against the background.

The possibility of deep cooling by CP was checked using data of
paper \cite{Baldin} in which a hint on existence of an ``excited
state of deutron" was noticed already in 1979. Our analysis showed
\cite{Kostenko2,Deuteron} that subtle contours of the dibaryon which
was observed by WASA-at-COSY Collaboration \cite{WASA} are seen
indeed in a right place suggested in \cite{Baldin}. Hereafter these
data are referred to as the Stavinskii group experiment (SGE) and
data from \cite{Troyan} as the Troyan experiment (TE) data.

The main surprise waited for us in the range of two other peaks
observed in  SGE, which were never  analyzed before but identified a
priori with elastic d-d and d-N scattering. Calculations revealed
this is definitely not so, and the peaks correspond very
surprisingly to transitions between different dibaryons from TE
\cite{Kostenko2,Deuteron}. It will be shown below that an
explanation of this unexpected result is connected with elucidation
of a transverse momentum anomaly found out in data of EVA group
\cite{EVA} in paper \cite{Kostenko3}. All the  facts  in the
aggregate will allow to assume a chance of experimental observation
of phase transitions in two-nucleon systems.

\section{Clarification from EVA experiment}
\label{sec-1}
We shall consider the total and relative momenta of intranuclear
proton (IP) and neutron (IN) affiliated to a short range correlation
(SRC), which are defined via momenta of IP and IN  as follows:
$$
\mathbf{p}^{cm}=\mathbf{p}_f + \mathbf{p}_n, \qquad
\mathbf{p}^{rel}= \mathbf{p}_f - \mathbf{p}_n .
$$
In the model of quasifree knockout (MQK),  $\mathbf{p}_f$  may be
expressed via momenta of the incoming proton and the secondary
registered ones in the following way \cite{EVA}:
$$
\mathbf{p}_f= \mathbf{p}_1 + \mathbf{p}_2 -\mathbf{p}_0 .
$$
It is also assumed in MQK that IN leaves the target nucleus C$^{12}$
without essential changes of the momentum $\mathbf{p}_n$ it had
before the interaction.

It was shown in \cite{EVA} that experimental data for longitudinal
(along $\mathbf{p}_0$) components of $\mathbf{p}^{cm}$ and
$\mathbf{p}^{rel}$ are in a good agreement with MQK and SRC. Our
analysis of the data confirmed this conclusion and gave the
following estimations for the values under consideration (hereafter
all momenta are in  GeV/c):
\begin{equation}\label{DataZ}
\left\langle {\mathrm{p}_z^{cm}} \right\rangle  \approx 0, \qquad
\sigma _z^{cm} \approx 0.1, \qquad \left\langle {\mathrm{p}_z^{rel}}
\right\rangle \approx 0.3, \qquad \sigma _z^{rel} \approx 0.1.
\end{equation}
Some additional work  allowed us to obtain also similar estimations
for the vertical, in the laboratory system, components of momenta
and they turned out to be very different (see \cite{Kostenko3}
for more details):
\begin{equation}\label{DataX}
\left\langle {\mathrm{p}_x^{cm}} \right\rangle  \approx 0, \qquad
\sigma _x^{cm} \approx 0.6, \qquad \left\langle {\mathrm{p}_x^{rel}}
\right\rangle \approx 0.6, \qquad \sigma _x^{rel} \approx 0.2.
\end{equation}
Mathematical modeling in the frame of the intranuclear cascade model
\cite{Barashenkov} did not reveal any visible influence of
intranuclear scattering (see \cite{Kostenko3})  and then interaction
between IP and IN was studied. It may be shown that in this case the
formulas which were used in \cite{EVA} for calculation of
$\mathbf{p}^{cm}$ and $\mathbf{p}^{rel}$ by means of the external
momenta $\mathbf{p}_0$, $\mathbf{p}_1$ and $\mathbf{p}_2$ give
\begin{equation}\label{pnew}
\mathbf{p}^{cm}=\mathbf{p}_f + \mathbf{p}_n, \qquad
\mathbf{p}^{rel}= \mathbf{p}_f - \mathbf{p}_n + 2\Delta
\mathbf{p}_f,
\end{equation}
where $\Delta \mathbf{p}_f$ is a momentum transfer from IN to IP.
Thus, the difference between $\left\langle {\mathrm{p}_z^{rel}}
\right\rangle$ and $\left\langle {\mathrm{p}_x^{rel}} \right\rangle$
may be, in principle, explained by the IP-IN interaction which
became involved.

A reason for the difference between $\sigma _z^{cm}$ and $\sigma
_x^{cm}$ is not so trivial. Whereas $\sigma _z^{cm}$ is well
explained by the intranuclear Fermi motion \cite{Kostenko3}, large
value of $\sigma _x^{cm}$ is provided by an unobvious quantum effect
firstly noted by L.D.~Landau and R.~Peierls \cite{Landau1, Landau2}. The fact
is that an attempt to measure $p_x^{cm}$ in the EVA experiment
exactly coincides with the well-known Landau-Peierls gedanken
experiment which demonstrates an inevitable influence of a momentum
measurement procedure on a final value of momentum if the process
lasts a limited time\footnote{A possible reason for finiteness of
the interaction time is the localization of projectile's wave
function in a restricted domain of space.} $\Delta t$. The
corresponding momentum perturbation of SRC may be written in an
explicit form as,
$$
\mathbf{p}^{cm}=(\mathbf{p}_f +\mathbf{\Delta}_f) + (\mathbf{p}_n
+\mathbf{\Delta}_n),
$$
where for a change of velocity of SRC, $\Delta
\mathbf{v}=(\mathbf{\Delta}_f+\mathbf{\Delta}_n)/(m_f +m_n)$, we
have $\Delta {v_z} = 0$, since $z$-component of the total momentum
is not measured, and
\begin{equation}\label{L-P}
\qquad \Delta {v_x} = \hbar /(\Delta {p_x}\Delta t).
\end{equation}
The last expression contains $\Delta {p_x}$ which is a precision of
measurement of $p_x^{cm}$. It may be estimated as $\sigma
_x^{rel}=\left\langle {\left| {{\Delta _{fx}} - {\Delta _{nx}}}
\right|} \right\rangle $. After substitution $\left\langle \Delta
{v_x} \right\rangle =\sigma _x^{cm}/(m_f +m_n) $ in (\ref{L-P}), we
find $ \Delta t  \sim 10^{-23}$ s and $ \Delta E  \sim \hbar/ \Delta
t  = 66$ MeV.

Now we can understand the reason for observation of excited deuteron
levels in the Stavinskii group experiment. Indeed, due to
Landau-Peierls uncertainty relation, masses of target and incident
deuterons in the initial state were not fixed exactly and some of
the excited dibaryon levels might show  themselves in the  peaks
observed. Overwhelming contribution of projectile's excited states
as compared with target's ones, which was established in
\cite{Kostenko2, Deuteron}, may be a manifestation of the
relativistic effect of time dilation that allows highly excited
states to exist much longer in the incident deuteron (see section
\ref{sec-4}).

Kinematics of EVA experiment was designed to select events with
$$
|(\mathbf{p}_f - \mathbf{p}_n)_x|<|(\mathbf{p}_f - \mathbf{p}_n)_z|
$$
due to a preferable choice of those of them in which IP was rapidly
moving forward in the same direction as the incident proton
(cross-section dependence on $\sqrt{s}$ described by the quark
counting rules \cite{Matveev}, \cite{Brodsky} was used). Taking into
account the last inequality and (\ref{DataZ}), (\ref{DataX}),
(\ref{pnew}), we obtain $2 \left\langle {\left| {\Delta {p_{fx}}}
\right|} \right\rangle > 0.3$. At the same time, validity of MQK for
z-direction, which was established by EVA collaboration \cite{EVA},
gives $\left\langle {\left| {\Delta {p_{fz}}} \right|} \right\rangle
\approx 0$. Significance and possible consequences of these two
relations are discussed in next section.

\section{Coherent dibaryons}
\label{sec-2} The most important observation concerning the results
of EVA is $2 \left\langle {\left| {\Delta {p_{fx}}} \right|}
\right\rangle > 0.3$ in combination with the quasifree knockout in
the longitudinal direction. This definitely implies  impossibility
to consider IP-IN interaction as elastic scattering initiated by
previous elastic projectile-IP interaction. Indeed, in a case like
that IP would obtain a recoil momentum directed in the transversal
plane and thus should necessarily transfer to IN some longitudinal
momentum. But experimentally we see something strictly opposite: the
whole of momentum transfer is confined to the transversal plane.
Therefore IP-IN interaction should propagate through a very unusual
intermediate state which has such a property. This conclusion is in
a qualitative agreement with the observation following from TE and
SGE about dimensionality of the 6-q excited
oscillator\footnote{Indeed, energy of the ground state of the
oscillator should be equal to $\hbar \omega /2 + \hbar \omega /2$
according to TE and SGE. Thus it might consist of one degree of
freedom oscillating in 2-D space or of two independent 1-D
oscillators. The ground state was observed as dibaryon with mass
1.886 GeV/c$^2$ in \cite{Troyan} and may also be extracted from
\cite{Baldin}  as a particle X in processes:  X+d $\to$ Y+d, d+X
$\to$ d+d, d+X $\to$ X+d,  X+X $\to$ X+d,  X+X $\to$ Y+d (see
\cite{Kostenko2,Deuteron}).}. We shall consider intermediate states
appearing in EVA experiment as 2-D coherent excitations of quantum
oscillator because they can maintain relative distance and momentum
of colliding particles with maximal accuracy permitted by the
uncertainty relation, $\Delta {p_i}(t)\Delta {x_i}(t) \sim \hbar$,
and transfer them safely in 2-D kinematics to outgoing particles,
thereby keeping their emission in the transversal plane. Thus
glueing two nucleons into a single 6-quark system together with
strong excitation of its inner oscillators seems to be the most
plausible explanation of the EVA experiment puzzle.

Before discussing Glauber's coherent state for 2-D oscillator, it is
convenient to define some vector notations \cite{Perelomov}:
$$\alpha = ({\alpha _1}, {\alpha _2}), \; a = ({a_1},{a_2}),
\; [n] = ({n_1},{n_2}), \;  [n]! = {n_1}!{n_2}!,
$$
$$
{({a^\dag })^{[n]}} =
{(a_1^\dag )^{{n_1}}}{(a_2^\dag )^{{n_2}}},\qquad \left| {[n]}
\right\rangle  = {({a^\dag })^{[n]}}\left| 0 \right\rangle /\sqrt
{[n]!} .
$$
The Glauber coherent state of 2-D oscillator may be
written as a unitary transformation of the ground one,
\begin{equation}\label{DGlaub}
\left| \alpha  \right\rangle  = D(\alpha )\left| 0 \right\rangle ,
\qquad D(\alpha ) = \exp (\alpha {a^\dag } - {\alpha ^*}a).
\end{equation}
This expression can be transformed to
\begin{equation}\label{2DGlauber}
\left| \alpha  \right\rangle  = \exp \left( {\alpha {a^\dag }}
\right)\left| 0 \right\rangle  = \exp \left( { - {{\left| \alpha
\right|}^2}/2} \right) \sum\limits_{[n]} {\frac{{{\alpha
^{[n]}}}}{{\sqrt {[n]!} }}} \left| {[n]} \right\rangle .
\end{equation}
Actually, (\ref{2DGlauber}) corresponds to two independent 1-D
oscillators swinging along $x_1$- and $x_2$-directions, and
$$
\left|
{[n]} \right\rangle  \equiv a_1^{\dag {n_1}}a_2^{\dag {n_2}}\left| 0
\right\rangle  \otimes \left| 0 \right\rangle /\sqrt {{n_1}!{n_2}!}
.
$$

It is also possible to define operators describing excitations with
negative and positive helicities,
$$
{a_ + } = ({a_1} + i{a_2})/\sqrt
2 , \;\; {a_ - } = ({a_1} - i{a_2})/\sqrt 2 ,
$$
accordingly. Corresponding basis vectors $\;\; \left| {[n]} \right\rangle =  a_+^{\dag
{n_+}}a_-^{\dag {n_-}} \left| 0 \right\rangle  \otimes \left| 0
\right\rangle / \sqrt {{n_+}!{n_-}!} $ are eigenvectors of operators
$${L_3} = {x_1}{p_2} - {p_1}{x_2} =\hbar(a_ - ^\dag {a_ - } - a_ +
^\dag {a_ + })=\hbar(n_- - n_+)$$ and $$ H = \hbar \omega (n_1+n_2 +
1)=\hbar \omega (n_+ +n_- +1).$$ Thus, it is possible to define
Glauber's coherent states with nonzero projections of orbital
momentum along $x_3$-axis of two different quarks belonging to the
same 6-q system.

Probabilities of registration of dibaryon lying on
$n$-th oscillator level are
$${w_n} = \exp ( - {\left| \alpha  \right|^2}) \sum\limits_{{n_i} +
{n_j} = n} {\frac{{{{\left| {{\alpha _i}} \right|}^{2{n_i}}}{{\left|
{{\alpha _j}} \right|}^{2{n_j}}}}}{{{n_i}!{n_j}!}}} , \qquad n=0, 1,
...  $$ Hereafter $i= 1$ or $+$ and $j= 2$ or $-$,
correspondingly.

It is  useful also to have explicit expressions for generalized
coherent states \cite{Perelomov} which may be observed
experimentally as coherent dibaryons too. They should be considered
as excitations above  different  ground states of oscillator, which
are  possible now and determined by different representations of
$su(1,1)$ algebra. In this case, quants of excitations are created
and annihilated in pairs and that is described by three generators
of $SU(1,1)$ group:
\begin{equation}\label{su}
K = {a_i}{a_j}, \qquad {K^\dag } = a_i^\dag a_j^\dag , \qquad {K_0}
= \frac{1}{2}\left( {a_i^\dag {a_i} + a_j^\dag {a_j} + 1} \right) ,
\end{equation}
 where \[ \left[ {{K_0},K} \right] =  - K, \qquad \left[
{{K_0}, {K^\dag }} \right] = {K^\dag }, \qquad  \left[ {K,{K^\dag }}
\right] = 2{K_0} .\] These coherent states are also defined as a
unitary transformation of one of the ground states,
$$
\left| \xi  \right\rangle  = D(\xi ){\left| 0 \right\rangle _k} = $$
\begin{equation}\label{GCS}
{\left( {1 - {{\left| \zeta \right|}^2}} \right)^k}\exp \left(
{\zeta {K^\dag }} \right){\left| 0 \right\rangle _k} ={\left( {1 -
{{\left| \zeta  \right|}^2}} \right)^k}{\sum\limits_{m = 0}^\infty
{\left[ {\frac{{\Gamma (m + 2k)}}{{m!\Gamma (2k)}}} \right]} ^{1/2}
}{\zeta ^m}\left| {k,k + m} \right\rangle ,
\end{equation}
\begin{equation}\label{DGen}
D(\xi ) = \exp (\xi {K^\dag } - {\xi ^*}K),
\end{equation}
index $k=1,3/2,2,5/2, ...$ determines a ground oscillator state,
\begin{equation}\label{param}
\zeta  = \tanh \left( {\left| \xi \right|} \right)\exp (i\psi ), \;
\beta = 2\ln \cosh \left( {\left| \xi  \right|} \right) =  - \ln
\left( {1 - {{\left| \zeta \right|}^2}} \right), \; \gamma  =  -
{\zeta ^*}.
\end{equation}

Energies of oscillator states are eigenvalues of the Hamiltonian,
\begin{equation}\label{Ham}
H\left| {k,k + m} \right\rangle  \equiv 2 \hbar \omega {K_0}\left|
{k,k + m} \right\rangle  = 2 \hbar \omega (k + m)\left| {k,k + m}
\right\rangle  , \;\; m=0,1, ...
\end{equation}
and the probabilities of the dibaryon observation are
$${w_m} = {\left( {1 -
{{\left| \zeta  \right|}^2}} \right)^{2k}}\frac{{\Gamma (m + 2k)}}
{{m!\Gamma (2k)}}{\left| \zeta  \right|^{2m}} .$$
Taking into
account \cite{Troyan, Baldin},  we should choose $k=1$. Then
$$
{w_m} ={\left( {1 - {{\left| \zeta  \right|}^2}} \right)^2}(m +
1){\left| \zeta \right|^{2m}},
$$
where  $\left| \zeta \right|<1$, see (\ref{param}).

\section{Coherent dibaryons and phase transitions}
\label{sec-3} There is a natural correlation between coherent states
and phase transitions, which becomes apparent if we shall study
metamorphoses of the oscillator Hamiltonian under unitary
transformations $ D(\alpha )$ and $ D(\xi )$ defined above by
(\ref{DGlaub}) and (\ref{DGen}). Good explanatory comments
concerning calculations in this section can be found in
\cite{Bogoliubov}.

Transformation of creation and annihilation operators corresponding
to (\ref{DGlaub})  is the following (in vector notations): $${a^\dag
} \to {a'^\dag } = {D^\dag }(\alpha ){a^\dag }D(\alpha ) = {a^\dag }
+ {\alpha ^*} , \qquad a \to a' = {D^\dag }(\alpha )aD(\alpha ) = a
+ \alpha .$$ This alteration leads to a change of the oscillator
Hamiltonian,
\[H = \hbar \omega ({a^\dag }a + 1) \to H' = \hbar \omega ({{a'}^\dag }a' + 1) =
\hbar \omega ({a^\dag }a + 1) + \hbar \omega \left[ {{{\left| \alpha
\right|}^2} + ({a^\dag }\alpha + a{\alpha ^*})} \right] .\] Thus we
see that the Hamiltonian gains two additional terms $\hbar \omega
{{\left| \alpha \right|}^2}$ and $\hbar \omega({a^\dag }\alpha +
a{\alpha ^*}). $ The first of them may be interpreted as energy
spent on creation of a complex field $\alpha$ (quantum condensate),
the second one describes an interaction between oscillator
excitations and this new field. In the general case, energy of the
system  varies due to the transformation (\ref{DGlaub}) which
therefore may be interpreted as describing a phase transition of the
first order.

Unitary transformation ${a^\dag } \to {a'^\dag } = {D^\dag }(\xi
){a^\dag }D(\xi )$ and $ a \to a' = {D^\dag }(\xi )a D(\xi  )$ with
$D(\xi )$ described by (\ref{DGen}) gives the Bogoliubov
transformation for Bose operators \cite{Bogoliubov1},
\begin{equation}\label{Bogol}
\begin{array}{l}
{a_i} \to {a_i}^\prime  = {a_i}\cosh \xi  + a_j^\dag \sinh \xi ,
\qquad a_i^\dag  \to {({a_i}^\prime )^\dag } = a_i^\dag \cosh \xi +
{a_j}\sinh \xi ,\\ \hspace{0.1cm} {a_j} \to {a_j}^\prime  =
{a_j}\cosh \xi + a_i^\dag \sinh \xi , \qquad a_j^\dag  \to
{({a_j}^\prime )^\dag } = a_j^\dag \cosh \xi  + {a_i}\sinh \xi .
\end{array}
\end{equation}
Then the oscillator Hamiltonian  takes the form:
$$
H = \hbar \omega ({a^\dag }a + 1) \to H' = \hbar \omega \left[
{2\cosh (2\xi ){K_0} + \sinh (2\xi )\left( {{K^\dag } + K} \right)}
\right].
$$
In the limit  $\xi \to 0$, we have $ H' = 2\hbar \omega {K_0},$ in
line with (\ref{Ham}). Subject to this extreme condition, we can
identify ground state ${\left| 0 \right\rangle _{k = 1}}$ of
$su(1,1)$ oscillator with state $\left| 0 \right\rangle  \otimes
\left| 1 \right\rangle $ of the usual 2-D oscillator, since it
satisfies the necessary characteristic properties of ${\left| 0
\right\rangle _{k = 1}}$:
$$
K \left| 0 \right\rangle  \otimes \left|
1 \right\rangle =0, \qquad K_0 \left| 0 \right\rangle  \otimes \left| 1
\right\rangle = \left| 0 \right\rangle  \otimes \left| 1
\right\rangle.
$$

So far as oscillator excitations are produced now by
pairs, the spectrum doubles (phase transition of the second order).
In the general case, when $\xi \ne 0$, relations (\ref{Bogol})
describe a first-order phase transition. It ends with nontrivial
renormalization of spectrum and creation of a quantum condensate
$\varphi=\sinh (2\xi )$ which interacts with pairs of oscillator
excitations. This phase transition occurs after the minimal
excitation of one of the oscillators, $\left| 0 \right\rangle \to
{\left| 0 \right\rangle _{k = 1}}$. Experiments \cite{Troyan,
Baldin} allow existence of these processes with small values of
$\varphi$.

There is a difference between phase transitions in macro- and
microsystems. An amplitude $\alpha$ of coherent state $ \left|
\alpha \right\rangle$ may be interpreted as a macroscopic field only
in the limit of large average number of excitations. Otherwise it
seems that a possibility of observing superpositions of coherent
states with different amplitudes may exist\footnote{One special case
of superposition of $N$ Glauber's coherent states was considered in
\cite{Birula}.}. Then for Glauber's coherent states, we should take
instead of (\ref{2DGlauber}):
$$
\left| C_i,C_j \right\rangle = \int
{d{\mu _i}} d{\mu _j}{C_i}({\mu _i}){C_j}({\mu _j})\exp (\chi {\mu
_i}a_i^\dag + \chi {\mu _j}a_j^\dag  - \mu _i^2/2 - \mu
_j^2/2\left)| 0 \right\rangle,
$$
where ${C_i}({\mu _i})$ and ${C_j}({\mu _j})$ describe the superposition
supposed, and $\alpha=\chi {\mu _i}$, $\chi=i$ or $-1$, $\mu$ is
a real number. It is obvious,
$$
\left| C_i,C_j \right\rangle ={{\tilde f}_i}(a_i^\dag ){{\tilde
f}_j}(a_j^\dag )\left| 0 \right\rangle ,
$$
where ${\tilde f}_i$ is Fourier or Laplace transform of
${C_i}({\mu _i})\exp ( - \mu_i^2/2).$
The generalized coherent states can be modified similarly
using (\ref{GCS}).

\section{Discussion and conclusion}
\label{sec-4}In this paper we have inverted the conventional logic
of utilization of canonical transformations in theory of phase
transitions. Usually they are applied for diagonalization of a model
Hamiltonian and estimation of its spectrum. Here we assumed that the
oscillator spectrum and coherent states were indeed disclosed in
\cite{Troyan, Baldin, EVA} for search of non-diagonalized
Hamiltonian which might generate the states hypothetically detected.

Although none of the discussed experimental papers \cite{Troyan,
Baldin, EVA} contains sufficient grounds for the definite conclusion
about existence of oscillator excitations in 6-quark systems,
interconnections between them appear striking. For example, let us
estimate numbers of levels observed in \cite{Baldin} using the
measurement time determined in \cite{EVA}. We can hope for nearly
the same value of the measurement time, though the measurement of
momentum of p-n pair is performed\footnote{For a quantum measurement
of the required type to be carried out it is sufficient  that an
information about stopping power of secondary particles to be stored
in an environment. A person may neither participate in making
measuring apparatus nor in observing results. According to
R.~Feynman, it is quite enough that {\it ``nature knows"}
\cite{Feynman}. } by scattering accelerated deuteron in
\cite{Baldin} rather than proton in \cite{EVA}. Then the
Landau-Peierls uncertainty relation tells us that we can observe 6-7
oscillator levels in target deuteron and, due to the relativistic
effect of time dilation, 32  levels in incident one. In
\cite{Kostenko2,Deuteron}, there were found 31 consecutive levels of
the second type in agreement with this estimation and only 2 of the
first type. It seems that the residual 4-5 of them were lost because
of the SRC-d scattering dominance over SRC-SRC one at least at the
experimental conditions in \cite{Baldin}, where the recoil deuterons
were registered.

An alternative explanation of the number of the levels observed in
\cite{Baldin} is identification of $10^{-23}$ s with a time of the
excitation energy transfer through SRC with subsequent repeating the
previous estimations. It is worth noting in this connection that
data obtaining during JLab Hall A experiment \cite{JLab} can
convince us even more easily than the EVA data in existence of this
energy transfer. Indeed, it is possible to check that without this
assumption a situation shown in Fig.~1 of \cite{JLab} corresponds to
violation of the energy conservation law at a level of 294 MeV.

Relying on the facts discussed above, it is reasonable to assume
that phase transitions of baryon matter can be observed at light
nuclei interactions as well as at heavy ions collisions usually
considered. As far as this important assumption is  based on the
results of only 3 experimental papers \cite{Troyan, Baldin, EVA},
they should be verified once again with an accuracy maximum possible
to enhance  reliability of the assertion. In any case, the surest
method to recognize what \emph{nature really knows} is to ask nature
itself about that.

The  assumption that the Landau-Peierls phenomenon has something to
do with an explanation of the EVA experiment \cite{EVA} is the most
unexpected one in this paper. It has nevertheless a chance of
success, though nothing like that was observed by the JLab Hall A
Collaboration \cite{JLab}, where a more convincing method of finding
$\mathbf{p}^{cm}$ was used. The fact is that beams with very
different wave packets of projectiles were employed in \cite{EVA}
and \cite{Hen}. According to \cite{Wohl} and \cite{Reece}, $(\Delta
p/p)_{EVA}=5$ $10^{-2}$ and $(\Delta E/E)_{Hall A}=10^{-4}$,
accordingly. This means that localization in space $\Delta x \sim
\hbar/\Delta p$ of wave packets of projectiles was different in
these experiments and thus the time of Landau-Peierls measurements
might be different. Using well-known properties of the Fourier
transformation, it is easy to check that the localization of
projectile electrons in JLab experiments was large than 214 fm.
Concerning the BNL beam for EVA we can only say that it might be
more than two order less, though a more exact estimation require
solving a difficult quantum state tomography problem.

In one of a recent review \cite{Hen}, representatives of the EVA and
JLab Hall A Collaborations arranged to take for granted that the
total momentum of the intranuclear pair observed by EVA conforms to
that observed by JLab. Our analysis revealed that this assumption
leads to a distinct worse agreement between simulated events and
data of the EVA experiment, see \cite{Kostenko3}. However, it is
more important that even in that case we still should agree with the
conclusion about exactly the same values of momentum transfer
through SRC pair in the transverse plane (see \cite{Kostenko3} for
more details). Existence of this momentum transfer (coupled with
quasi-free knock-out in the longitudinal direction observed in
\cite{EVA}) is the main experimental evidence for the conclusion
about the possibility of the coherent dibaryon excitation inside n-p
SRC.

The review \cite{Hen} together with a recent publication \cite{CLAS}
suggests an important development of the previous version of the SRC
model due to L.~Frankfurt and M.~Strikman \cite{Frankfurt}. This
change  requires comparison with a consideration in the present
paper and deserves inserting the renovation into a general context
of already existing physics. The main assertion in the new version
of SRC model is a recognition of the fact that ``nucleons are
modified substantially when they fluctuate into SRC pairs"
\cite{CLAS}. Authors of \cite{Hen} rename the previous version of
the SRC correlations into 2N-SRC ones, or two-nucleon clusters, and
reserve the name SRC correlations for the new hypothetical objects.
The renaming like that is hardly reasonable and may lead to a
confusion. Logically, it would be better to save the old name for
the Frankfurt-Strikman SRC and to give a new denomination to the new
objects. Since the suggested fluctuations of the nuclear density are
supposed to have modified quark momentum distribution which is the
characteristic property of the fluctons
\cite{Blokh,Efremov,BLT1,LT,Efremov1,BLT2}, that makes sense to
designate them as the {\it flucton phase} of the SRC pairs.

It is also important to note that the new ground state fluctuations
in nuclear matter were reported to be found in neutron-proton SRC
pairs only, see \cite{CLAS}. According to the present paper, the
coherent 6-q states which might be observed  in the EVA experiment
were excited in n-p SRC pairs too, after their interactions with the
projectile protons. This gives a hint that the coherent dibaryons
are {\it excited} states of the {\it ground} state fluctuations
suggested by the CLAS Collaboration. Besides, quarks oscillator
levels, existence of which are the necessary condition for existence
of the coherent dibaryon, were presumably observed by Baldin et al.
\cite{Baldin} also in n-p systems after hard scattering the
projectile off it. The same levels as in \cite{Baldin} were also
observed after ``deep cooling" of highly excited n-p systems by
secondary pions in \cite{Troyan}. That all leads us to one more
possible explanation of the difference between the results of the
EVA and JLab Hall A collaborations experiments different from the
Landau-Peierls mechanism suggested above. Indeed, it is rational to
assume that the flucton phase which is allowed only in
neutron-proton SRC is characterized by a more intense interaction
with the residual nucleus than the usual SRC and therefore by a
wider momentum distribution in the transverse plane. Now such an
explanation can be considered as alternative to the Landau-Peierls
mechanism.

In conclusion, investigation of SRC persists to be an actual branch
of physics containing many unsolved puzzles. Recent developments
show that it also opens the door into the area of phase transitions
in hadronic matter. Interrelations found in the present paper
inspire exploration of the n-p systems first of all to obtain
information about the 6-q system in its ground and excited states.

\section*{Acknowledgment}
I thank my co-authors J. Pribi\v{s} and V.~Filinova which assisted
me with numerical calculations in an early stage of the study.
Valuable remarks of S.S.~Shimanskiy are gratefully acknowledged. I
dedicate this paper to \fbox{M.Z.~Yuriev} for many valuable
discussions.

\end{document}